\RequirePackage{ifpdf}
\ifpdf 
\documentclass[pdftex]{sigma}
\else
\documentclass{sigma}
\fi

\begin{document}

\allowdisplaybreaks

\renewcommand{\thefootnote}{$\star$}

\renewcommand{\PaperNumber}{074}

\FirstPageHeading

\ShortArticleName{A Class of Special Solutions for the Ultradiscrete Painlev\'e II Equation}

\ArticleName{A Class of Special Solutions \\
for the Ultradiscrete Painlev\'e II Equation\footnote{This
paper is a contribution to the Proceedings of the Conference ``Integrable Systems and Geomet\-ry'' (August 12--17, 2010, Pondicherry University, Puducherry, India). The full collection is available at \href{http://www.emis.de/journals/SIGMA/ISG2010.html}{http://www.emis.de/journals/SIGMA/ISG2010.html}}}

\Author{Shin ISOJIMA and Junkichi SATSUMA}

\AuthorNameForHeading{S.~Isojima and J.~Satsuma}

\Address{Department of Physics and Mathematics, Aoyama Gakuin University,\\
5-10-1 Fuchinobe, Chuo-ku, Sagamihara-shi, Kanagawa, 252-5258, Japan}

\Email{\href{mailto:isojima@gem.aoyama.ac.jp}{isojima@gem.aoyama.ac.jp}, \href{mailto:satsuma@gem.aoyama.ac.jp}{satsuma@gem.aoyama.ac.jp}}

\ArticleDates{Received April 01, 2011, in f\/inal form July 14, 2011;  Published online July 22, 2011}

\Abstract{A class of special solutions are constructed in an intuitive way for the ultradiscrete analog of $q$-Painlev\'{e} II ($q$-PII) equation. The solutions are classif\/ied into four groups depending on the function-type and the system parameter.}


\Classification{34M55; 33E30; 39A13}

\section{Introduction}
Ultradiscretization \cite{TTMS} is a limiting procedure transforming a given dif\/ference equation into a cellular automaton, in which dependent variables also take discrete values. To apply this procedure, we f\/irst replace a dependent variable $x_n$ in the equation by
\begin{equation}
x_n = e^{X_n / \varepsilon}, \label{eq:exp_replace}
\end{equation}
where $\varepsilon$ is a positive parameter. Next, we apply $\varepsilon \log$ to both sides of the equation and take the limit $\varepsilon \to +0$. Then, using identity
\begin{equation*}
\lim_{\varepsilon \to +0} \varepsilon \log \big(e^{X / \varepsilon} + e^{Y / \varepsilon}\big) = \max (X,Y),
\end{equation*}
the original dif\/ference equation is approximated by a piecewise linear equation which can be regarded as a time evolution rule for a cellular automaton. In many examples, cellular automata obtained by this systematic method preserve the essential properties of the original equations, such as the qualitative behavior of exact solutions. However, the ansatz \eqref{eq:exp_replace} is only possible if the variable $x_n$ is positive def\/inite. This restriction is called `negative problem'.

From theoretical and application points of view, it is an interesting problem to study ultradiscrete analogs of special functions and their def\/ining equations, including the Painlev\'e equations. Ultradiscrete analogs for some of the Painlev\'e equations and their special solutions are discussed, for example, in \cite{GORTT,TTGOR,RTGO}. However, the class of solutions for ultradiscrete Painlev\'e equations has been restricted because of the negative problem. Some attempts resolving this problem are reported, for example, in \cite{IGRS, KL, Orm}. The authors and coworkers study in \cite{IGRS} an ultradiscrete Painlev\'e II equation with $\sinh$ ansatz and discuss its special solution of Bi function type.

In order to overcome the negative problem, a new method `ultradiscretization with parity variables' ($p$-ultradiscretization) is proposed in~\cite{MIMS}. The procedure keeps track of the sign of original variables. By using this method, the authors and coworkers present~\cite{IKMMS} a $p$-ultradiscrete analog of the $q$-Painlev\'e II equation ($q$-PII),
\begin{equation}
\left( z (q \tau) z (\tau) + 1 \right) \left( z (\tau) z \big(q^{-1} \tau\big) + 1 \right) = \frac{a \tau^2 z (\tau)}{\tau - z (\tau)}. \label{eq:qPII_z}
\end{equation}
In \cite{IKMMS}, we also discuss a series of special solutions corresponding to that of $q$-PII written in the determinants of size $N$. However, the resulting solutions are reduced to only one solution for the $p$-ultradiscrete Painlev\'e II (udPII) equation. In this paper, we construct other series of special solutions for udPII and discuss their structure. In Section~\ref{section2}, we introduce the results in \cite{IKMMS} for the $p$-ultradiscrete Airy equation. Then, we construct special solutions for udPII in Section~\ref{section3}. These solutions are, from their construction, considered to be counterparts of those of $q$-PII written by the determinants. Finally, concluding remarks are given in Section~\ref{section4}.

\section{Ultradiscrete Airy equation with parity variables}\label{section2}

We start with a $q$-dif\/ference analog of the Airy equation
\begin{equation}
w (q \tau) - \tau w (\tau) + w \big(q^{-1} \tau\big) = 0, \label{eq:q-Airy}
\end{equation}
which reduces to the Airy equation
\begin{equation*}
\frac{d^2 v}{d s^2} + s v = 0
\end{equation*}
in a continuous limit.

In order to ultradiscretize \eqref{eq:q-Airy}, we put $\tau = q^m$ and $q = e^{Q / \varepsilon}$ ($Q<0$). Furthermore, we introduce an ansatz for $p$-ultradiscretization,
\begin{equation*}
w (q^m) = \{s (\omega_m) - s (-\omega_m)\} e^{W_m / \varepsilon}, 
\end{equation*}
where $\omega_m \in \{+1, -1 \}$ denotes the sign of $w (q^m)$ and $s (\omega)$ is def\/ined by
\begin{equation*}
s (\xi) =
\begin{cases}
1, & \xi = +1, \\
0, & \xi = -1.
\end{cases}
\end{equation*}
Taking the ultradicrete limit, we obtain a $p$-ultradiscrete analog of the Airy equation
\begin{gather}
\max \left(W_{m+1} + S (\omega_{m+1}),  mQ + W_{m} + S (-\omega_{m}),  W_{m-1} + S (\omega_{m-1}) \right) \nonumber\\
\qquad{} = \max \left(W_{m+1} + S (-\omega_{m+1}),  mQ + W_{m} + S (\omega_{m}),  W_{m-1} + S (-\omega_{m-1}) \right), \label{eq:ud-Airy}
\end{gather}
where $S (\omega)$ is def\/ined by
\begin{equation*}
S (\omega) =
\begin{cases}
0, & \omega = +1, \\
- \infty, & \omega = -1.
\end{cases}
\end{equation*}
An ultradiscretized variable is represented by a pair of $\omega_m$ and $W_m$, which is denoted as $\mathcal{W}_n = (\omega_m, W_m)$ in what follows. It is possible to rewrite the implicit form~\eqref{eq:ud-Airy} into explicit forward schemes
\begin{gather*}
\omega_{m+1}  =
\begin{cases}
\dfrac{\omega_{m} - \omega_{m-1}}{2} + \dfrac{\omega_{m} + \omega_{m-1}}{2} \mathrm{sgn} (F_{m}), & \omega_{m} = -\omega_{m-1} \text{ or } F_{m} \neq 0, \vspace{1mm}\\ 
\text{indef\/inite}, & \omega_{m} = \omega_{m-1} \text{ and } F_{m} = 0,
\end{cases} \\
W_{m+1} \;
 \begin{cases}
= \max \left(mQ + W_{m},  W_{m-1}\right),  &  \omega_{m} = -\omega_{m-1} \text{ or } F_{m} \neq 0, \\
\le W_{m-1}, & \omega_{m} = \omega_{m-1} \text{ and } F_{m} = 0,
\end{cases}
\end{gather*}
where $F_{m} := mQ + W_{m} - W_{m-1}$. Note that we generally have both of unique and indeterminate schemes depending on given values of $(\omega_{m}, W_{m})$ and $(\omega_{m-1}, W_{m-1})$. The explicit backward schemes are obtained by replacing~$m \pm 1$ with~$m \mp 1$, respectively.

We f\/ind two typical solutions of \eqref{eq:ud-Airy}. One is an Ai-function-type solution for $\mathcal{W}_0 = (+1, 0)$ and $\mathcal{W}_1 = (+1, 0)$,
\begin{gather*}
\mathrm{uAi} (m)  = (\omega_m, W_m) =
\begin{cases}
\left((-1)^{\frac{m(m-1)}{2}}, 0 \right),  & m \ge 0, \vspace{1mm}\\
\left(+1, \dfrac{m(m-1)}{2}Q \right), & m \le -1,
\end{cases}
\end{gather*}
and the other is a Bi-function-type for $\mathcal{W}_0 = (+1, 0)$ and $\mathcal{W}_1 = (-1, 0)$,
\begin{gather*}
\mathrm{uBi} (m) = (\omega_m, W_m) =
\begin{cases}
\left((-1)^{\frac{m (m+1)}{2}},  0 \right), & m \ge 0, \vspace{1mm}\\
\left(+1,-\dfrac{m(m+1)}{2}Q \right), & m \le -1.
\end{cases}
\end{gather*}
They show similar behavior as those of the Ai and Bi functions, respectively.

\section{Ultradiscrete Painlev\'{e} II equation with parity variables}\label{section3}

For the following discussion, we f\/irst introduce the results for \eqref{eq:qPII_z}.
It has been shown in \cite{HKW} that
\begin{equation}
z^{(N)} (\tau) =
\begin{cases}
\dfrac{g^{(N)} (\tau) g^{(N+1)} (q \tau)}{q^N g^{(N)} (q \tau) g^{(N+1)} (\tau)},
 & N \ge 0,\vspace{2mm}\\
\dfrac{g^{(N)} (\tau) g^{(N+1)} (q \tau)}{q^{N+1} g^{(N)} (q \tau) g^{(N+1)} (\tau)}, & N<0
\end{cases}
 \label{eq:gtoz}
\end{equation}
solves \eqref{eq:qPII_z} with $a=q^{2N+1}$, where the functions $g^{(N)} (t)$ ($N \in \mathbb{Z}$) satisfy the bilinear equations
\begin{gather}
 q^{2N} g^{(N+1)} (q^{-1} \tau) g^{(N)} (q^2 \tau) - q^{N} \tau g^{(N+1)} (\tau) g^{(N)} (q \tau) + g^{(N+1)} (q \tau) g^{(N)} (\tau) = 0, \label{eq:bilin1+}\\
 q^{2N} g^{(N+1)} (q^{-1} \tau) g^{(N)} (q \tau) - q^{2N} \tau g^{(N+1)} (\tau) g^{(N)} (\tau) + g^{(N+1)} (q \tau) g^{(N)} (q^{-1} \tau) = 0 \label{eq:bilin2+}
\end{gather}
 for $N \ge 0$ and
\begin{gather}
 q^{2N+2} g^{(N+1)} (q^{-1} \tau) g^{(N)} (q^2 \tau) - q^{N+1} \tau g^{(N+1)} (\tau) g^{(N)} (q \tau) + g^{(N+1)} (q \tau) g^{(N)} (\tau) = 0 , \label{eq:bilin1-} \\
 q^{2N+2} g^{(N+1)} (q^{-1} \tau) g^{(N)} (q \tau) - q^{2N+1} \tau g^{(N+1)} (\tau) g^{(N)} (\tau) + g^{(N+1)} (q \tau) g^{(N)} (q^{-1} \tau) = 0 \label{eq:bilin2-}
\end{gather}
for $N<0$. It is also known that $g^{(N)}(\tau)$ are written in terms of the Casorati determinant of size $|N|$ whose elements are represented by the solutions of \eqref{eq:q-Airy}.

In order to construct ultradiscrete analogs of these equations, we put $\tau = q^m$, $q = e^{Q / \varepsilon} (Q<0)$ and $a = e^{A / \varepsilon}$. Furthermore, we introduce
\begin{gather*}
z (q^m)  = (s (\zeta_m) - s (-\zeta_m)) e^{Z_m / \varepsilon}, \\
g^{(N)} (q^m)  = (s (\gamma_m^{(N)}) - s (-\gamma_m^{(N)})) e^{G_m^{(N)} / \varepsilon}. 
\end{gather*}
Then \eqref{eq:qPII_z} is reduced to udPII,
\begin{gather}
\max\Bigl[
Z_{m+1} + 3Z_{m} + Z_{m-1} +
\max \bigl\{  S (\zeta_{m+1}) + S (\zeta_{m}) + S (\zeta_{m-1}), \nonumber \\
\qquad S (-\zeta_{m+1}) + S (\zeta_{m}) + S (-\zeta_{m-1}),  S (-\zeta_{m+1}) + S (-\zeta_{m}) + S (\zeta_{m-1}), \nonumber \\
\qquad S (\zeta_{m+1}) + S (-\zeta_{m}) + S (-\zeta_{m-1}) \bigr\},  Z_{m+1} + 2Z_{m} + S (\zeta_{m+1}), \nonumber \\
\qquad 2Z_{m} + Z_{m-1} + S (\zeta_{m-1}), Z_{m} +  S (\zeta_{m}), Z_{m} + A + 2m Q + S (\zeta_{m}), \nonumber \\
\qquad Z_{m+1} + 2Z_{m} + Z_{m-1} + mQ +
\max \bigl\{  S (-\zeta_{m+1}) + S (\zeta_{m-1}), S (\zeta_{m+1}) + S (-\zeta_{m-1}) \bigr\}, \nonumber \\
\qquad Z_{m+1} + Z_{m} + mQ +
\max \bigl\{  S (-\zeta_{m+1}) + S (\zeta_{m}), S (\zeta_{m+1}) + S (-\zeta_{m}) \bigr\}, \nonumber \\
\qquad Z_{m} + Z_{m-1} + mQ +
\max \bigl\{  S (-\zeta_{m}) + S (\zeta_{m-1}), S (\zeta_{m}) + S (-\zeta_{m-1}) \bigr\} \Bigr] \nonumber \\
= \max\Bigl[
Z_{m+1} + 3Z_{m} + Z_{m-1} +
\max \bigl\{  S (-\zeta_{m+1}) + S (-\zeta_{m}) + S (-\zeta_{m-1}), \nonumber \\
\qquad S (\zeta_{m+1}) + S (-\zeta_{m}) + S (\zeta_{m-1}),  S (\zeta_{m+1}) + S (\zeta_{m}) + S (-\zeta_{m-1}), \nonumber \\
\qquad S (-\zeta_{m+1}) + S (\zeta_{m}) + S (\zeta_{m-1}) \bigr\},  Z_{m+1} + 2Z_{m} + S (-\zeta_{m+1}), \nonumber \\
\qquad 2Z_{m} + Z_{m-1} + S (-\zeta_{m-1}), Z_{m} +  S (-\zeta_{m}), Z_{m} + A + 2m Q + S (-\zeta_{m}), \nonumber \\
\qquad Z_{m+1} + 2Z_{m} + Z_{m-1} + mQ +
\max \bigl\{  S (\zeta_{m+1}) + S (\zeta_{m-1}), S (-\zeta_{m+1}) + S (-\zeta_{m-1}) \bigr\}, \nonumber \\
\qquad Z_{m+1} + Z_{m} + mQ +
\max \bigl\{  S (\zeta_{m+1}) + S (\zeta_{m}), S (-\zeta_{m+1}) + S (-\zeta_{m}) \bigr\}, \nonumber \\
\qquad Z_{m} + Z_{m-1} + mQ +
\max \bigl\{  S (\zeta_{m}) + S (\zeta_{m-1}), S (-\zeta_{m}) + S (-\zeta_{m-1}) \bigr\}, mQ \Bigr]. \label{eq:udP2}
\end{gather}
For \eqref{eq:bilin1+} and \eqref{eq:bilin2+}, we have their ultradiscrete analogs
\begin{gather}
\max \Bigl[  2NQ + G_{m-1}^{(N+1)} + G_{m+2}^{(N)}
 + \max \left\{ S \big(\gamma_{m-1}^{(N+1)}\big) + S \big(\gamma_{m+2}^{(N)}\big), S \big({-}\gamma_{m-1}^{(N+1)}\big) + S \big({-}\gamma_{m+2}^{(N)}\big) \right\}, \nonumber \\
 \qquad (N + m) Q + G_{m}^{(N+1)} + G_{m+1}^{(N)}  \nonumber\\
 \qquad{}  + \max \left\{ S \big(\gamma_{m}^{(N+1)}\big) + S \big({-}\gamma_{m+1}^{(N)}\big),
 S \big({-}\gamma_{m}^{(N+1)}\big) + S \big(\gamma_{m+1}^{(N)}\big) \right\}, \nonumber \\
\qquad G_{m+1}^{(N+1)} + G_{m}^{(N)} + \max \left\{ S \big(\gamma_{m+1}^{(N+1)}\big) + S \big(\gamma_{m}^{(N)}\big), S \big({-}\gamma_{m+1}^{(N+1)}\big) + S \big({-}\gamma_{m}^{(N)}\big) \right\} \Bigr] \nonumber \\
= \max \Bigl[  2NQ\! + G_{m-1}^{(N+1)}\! + G_{m+2}^{(N)}\!
 + \max \left\{ S \big(\gamma_{m-1}^{(N+1)}\big)\! + S \big({-}\gamma_{m+2}^{(N)}\big), S \big({-}\gamma_{m-1}^{(N+1)}\big)\! + S \big(\gamma_{m+2}^{(N)}\big) \right\}, \nonumber \\
\qquad (N + m) Q + G_{m}^{(N+1)} + G_{m+1}^{(N)} \nonumber \\
 \qquad {}+ \max \left\{ S \big(\gamma_{m}^{(N+1)}\big) + S \big(\gamma_{m+1}^{(N)}\big), S \big({-}\gamma_{m}^{(N+1)}\big) + S \big({-}\gamma_{m+1}^{(N)}\big) \right\}, \nonumber \\
\qquad G_{m+1}^{(N+1)} + G_{m}^{(N)} + \max \left\{ S \big(\gamma_{m+1}^{(N+1)}\big) + S \big({-}\gamma_{m}^{(N)}\big), S \big({-}\gamma_{m+1}^{(N+1)}\big) + S \big(\gamma_{m}^{(N)}\big) \right\} \Bigr] \label{eq:bilinearG1+imp}
\end{gather}
and
\begin{gather}
\max \Bigl[  2NQ + G_{m-1}^{(N+1)} + G_{m+1}^{(N)}
 + \max \left\{ S\big(\gamma_{m-1}^{(N+1)}\big) + S \big(\gamma_{m+1}^{(N)}\big), S \big({-}\gamma_{m-1}^{(N+1)}\big) + S \big({-}\gamma_{m+1}^{(N)}\big) \right\}, \nonumber \\
 \qquad (2N + m) Q + G_{m}^{(N+1)} + G_{m}^{(N)} \nonumber \\
 \qquad {} + \max \left\{ S \big(\gamma_{m}^{(N+1)}\big) + S \big({-}\gamma_{m}^{(N)}\big), S \big({-}\gamma_{m}^{(N+1)}\big) + S \big(\gamma_{m}^{(N)}\big) \right\}, \nonumber \\
 \qquad G_{m+1}^{(N+1)} + G_{m-1}^{(N)} + \max \left\{ S \big(\gamma_{m+1}^{(N+1)}\big) + S (\gamma_{m-1}^{(N)}\big), S \big({-}\gamma_{m+1}^{(N+1)}\big) + S \big({-}\gamma_{m-1}^{(N)}\big) \right\} \Bigr] \nonumber \\
= \max \Bigl[  2NQ + G_{m-1}^{(N+1)} + G_{m+1}^{(N)} \nonumber \\
 \qquad{} + \max \left\{ S \big(\gamma_{m-1}^{(N+1)}\big) + S \big({-}\gamma_{m+1}^{(N)}\big), S \big({-}\gamma_{m-1}^{(N+1)}\big) + S \big(\gamma_{m+1}^{(N)}\big) \right\}, \nonumber \\
\qquad (2N + m) Q + G_{m}^{(N+1)} + G_{m}^{(N)} \nonumber \\
\qquad{} + \max \left\{ S \big(\gamma_{m}^{(N+1)}\big) + S \big(\gamma_{m}^{(N)}\big), S \big({-}\gamma_{m}^{(N+1)}\big) + S \big({-}\gamma_{m}^{(N)}\big) \right\}, \nonumber \\
\qquad G_{m+1}^{(N+1)} + G_{m-1}^{(N)} + \max \left\{ S \big(\gamma_{m+1}^{(N+1)}\big) + S \big({-}\gamma_{m-1}^{(N)}\big), S \big({-}\gamma_{m+1}^{(N+1)}\big) + S \big(\gamma_{m-1}^{(N)}\big) \right\} \Bigr], \label{eq:bilinearG2+imp}
\end{gather}
respectively. For  \eqref{eq:bilin1-} and \eqref{eq:bilin2-}, we have
\begin{gather}
\max \Bigl[  2(N+1) Q + G_{m-1}^{(N+1)} + G_{m+2}^{(N)} \nonumber\\
\qquad{}
 + \max \left\{ S \big(\gamma_{m-1}^{(N+1)}\big) + S \big(\gamma_{m+2}^{(N)}\big), S \big({-}\gamma_{m-1}^{(N+1)}\big) + S \big({-}\gamma_{m+2}^{(N)}\big) \right\}, \nonumber \\
\qquad{} (N+m+1) Q + G_{m}^{(N+1)} + G_{m+1}^{(N)} \nonumber \\
\qquad{} + \max \left\{ S \big(\gamma_{m}^{(N+1)}\big) + S \big({-}\gamma_{m+1}^{(N)}\big), S \big({-}\gamma_{m}^{(N+1)}\big)
+ S \big(\gamma_{m+1}^{(N)}\big) \right\}, \nonumber \\
\qquad G_{m+1}^{(N+1)} + G_{m}^{(N)} + \max \left\{ S \big(\gamma_{m+1}^{(N+1)}\big) + S (\gamma_{m}^{(N)}\big), S \big({-}\gamma_{m+1}^{(N+1)}\big) + S \big({-}\gamma_{m}^{(N)}\big) \right\} \Bigr] \nonumber \\
= \max \Bigl[  2(N+1) Q + G_{m-1}^{(N+1)} + G_{m+2}^{(N)} \nonumber \\
 \qquad{} + \max \left\{ S \big(\gamma_{m-1}^{(N+1)}\big) + S \big({-}\gamma_{m+2}^{(N)}\big), S \big({-}\gamma_{m-1}^{(N+1)}\big) + S \big(\gamma_{m+2}^{(N)}\big) \right\}, \nonumber \\
\qquad (N+m+1) Q + G_{m}^{(N+1)} + G_{m+1}^{(N)} \nonumber \\
\qquad{} + \max \left\{ S \big(\gamma_{m}^{(N+1)}\big) + S \big(\gamma_{m+1}^{(N)}\big), S \big({-}\gamma_{m}^{(N+1)}\big) + S \big({-}\gamma_{m+1}^{(N)}\big) \right\}, \nonumber \\
\qquad G_{m+1}^{(N+1)} + G_{m}^{(N)} + \max \left\{ S \big(\gamma_{m+1}^{(N+1)}\big) + S \big({-}\gamma_{m}^{(N)}\big), S \big({-}\gamma_{m+1}^{(N+1)}\big) + S \big(\gamma_{m}^{(N)}\big) \right\} \Bigr] \label{eq:bilinearG1-imp}
\end{gather}
and
\begin{gather}
\max \Bigl[ 2(N+1) Q + G_{m-1}^{(N+1)} + G_{m+1}^{(N)} \nonumber \\
+ \max \left\{ S \big(\gamma_{m-1}^{(N+1)}\big) + S \big(\gamma_{m+1}^{(N)}\big), S \big({-}\gamma_{m-1}^{(N+1)}\big) + S \big({-}\gamma_{m+1}^{(N)}\big) \right\}, \nonumber \\
\quad (2N+m+1) Q + G_{m}^{(N+1)} + G_{m}^{(N)} \nonumber \\
\qquad{}+ \max \left\{ S \big(\gamma_{m}^{(N+1)}\big) + S \big({-}\gamma_{m}^{(N)}\big), S \big({-}\gamma_{m}^{(N+1)}\big) + S \big(\gamma_{m}^{(N)}\big) \right\}, \nonumber \\
\qquad G_{m+1}^{(N+1)} + G_{m-1}^{(N)} + \max \left\{ S \big(\gamma_{m+1}^{(N+1)}\big) + S \big(\gamma_{m-1}^{(N)}\big), S \big({-}\gamma_{m+1}^{(N+1)}\big) + S \big({-}\gamma_{m-1}^{(N)}\big) \right\} \Bigr] \nonumber \\
= \max \Bigl[2(N+1) Q + G_{m-1}^{(N+1)} + G_{m+1}^{(N)} \nonumber \\
\qquad{} + \max \left\{ S \big(\gamma_{m-1}^{(N+1)}\big) + S \big({-}\gamma_{m+1}^{(N)}\big), S \big({-}\gamma_{m-1}^{(N+1)}\big) + S \big(\gamma_{m+1}^{(N)}\big) \right\}, \nonumber \\
\qquad (2N+m+1) Q + G_{m}^{(N+1)} + G_{m}^{(N)} \nonumber \\
\qquad + \max \left\{ S \big(\gamma_{m}^{(N+1)}\big) + S \big(\gamma_{m}^{(N)}\big), S \big({-}\gamma_{m}^{(N+1)}\big) + S \big({-}\gamma_{m}^{(N)}\big) \right\}, \nonumber \\
\qquad G_{m+1}^{(N+1)} + G_{m-1}^{(N)} + \max \left\{ S \big(\gamma_{m+1}^{(N+1)}\big) + S \big({-}\gamma_{m-1}^{(N)}\big), S \big({-}\gamma_{m+1}^{(N+1)}\big) + S \big(\gamma_{m-1}^{(N)}\big) \right\} \Bigr], \label{eq:bilinearG2-imp}
\end{gather}
respectively. Finally, the transformations \eqref{eq:gtoz} are reduced to
\begin{gather}
\zeta_m^{(N)}  = \gamma_{m}^{(N)} \gamma_{m+1}^{(N+1)} \gamma_{m}^{(N+1)} \gamma_{m+1}^{(N)}, \label{eq:udCH1+} \\
Z_{m}^{(N)} = G_{m}^{(N)} + G_{m+1}^{(N+1)} - G_{m}^{(N+1)} - G_{m+1}^{(N)} - NQ \label{eq:udCH2+}
\end{gather}
for $N \ge 0$ and
\begin{gather}
\zeta_m^{(N)}  = \gamma_{m}^{(N)} \gamma_{m+1}^{(N+1)} \gamma_{m}^{(N+1)} \gamma_{m+1}^{(N)}, \label{eq:udCH1-} \\
Z_{m}^{(N)} = G_{m}^{(N)} + G_{m+1}^{(N+1)} - G_{m}^{(N+1)} - G_{m+1}^{(N)} - (N+1) Q \label{eq:udCH2-}
\end{gather}
for $N < 0$. If we f\/ind solutions for the ultradiscrete bilinear equations, special solutions for udPII are obtained through \eqref{eq:udCH1+}--\eqref{eq:udCH2-}.

Hereafter we consider only the case of $A = (2N+1) Q$ in \eqref{eq:udP2}, which corresponds to $a = q^{2N+1}$ in the discrete system. Firstly, we present the results reported in~\cite{IKMMS}, that is, the Ai-function-type solutions for $N \ge 0$. Solutions of~\eqref{eq:bilinearG1-imp} and~\eqref{eq:bilinearG2-imp} are given by $\mathcal{G}_m^{(N)} = (\gamma_m^{(N)}, G_m^{(N)})$ for $N=0, 1, 2, \dots$, where
\begin{gather*}
\gamma_m^{(N)}  =
\begin{cases}
\gamma_0^{(N)}(-1)^{\frac{m(m-1)N}{2}}, & m \ge 0, \\
\gamma_0^{(N)}, & m \le -1,
\end{cases} \\
G_m^{(N)}  =
\begin{cases}
 \dfrac{mN(N-1)}{2}Q + G_0^{(N)}, & m \ge 0,\vspace{2mm}\\
\dfrac{mN(m+N-2)}{2}Q + G_0^{(N)}, & m \le -1.
\end{cases} 
\end{gather*}
Since $G_m^{(N)} \to -\infty$ as $m \to -\infty$ in the same way as the uAi function, we call these solutions the Ai-function-type solutions.
From these solutions, we have only one special solution of udPII with $A=(2N+1)Q$ for $N=0, 1, 2, \dots$
\begin{gather}
\mathcal{Z}_m^{(N)} = \big(\zeta_m^{(N)}, Z_m^{(N)} \big) =
\begin{cases}
\left((-1)^m, 0 \right), & m \ge 0, \\
\left(+1, mQ \right), & m \le -1,
\end{cases} \label{eq:Psol_ai+}
\end{gather}
which does not depend on $N$. We note that $Z_m^{(N)} \to \infty$ as $m \to -\infty$.

Secondly, we investigate Bi-function-type solutions for $N \ge 0$. We f\/ind that $\mathcal{G}_{m}^{(0)} = (+1,0)$ and $\mathcal{G}_{m}^{(1)} = \mathrm{uBi} (m)$ solve \eqref{eq:bilinearG1+imp} and \eqref{eq:bilinearG2+imp} with $N=0$. By using this result, we inductively construct solutions $\mathcal{G}_{m}^{(N+1)}$ of the equations with $N \ge 1$ for a given function $\mathcal{G}_{m}^{(N)}$ and assigned values of $\mathcal{G}_{0}^{(N+1)}$ and $\mathcal{G}_{1}^{(N+1)}$. We further assume that $\mathcal{G}_{0}^{(N+1)}$ and $\mathcal{G}_{1}^{(N+1)}$ are chosen so that $\mathcal{G}_{m}^{(N+1)}$ for any $m$ are uniquely determined in \eqref{eq:bilinearG1+imp} and \eqref{eq:bilinearG2+imp}. Then we have the following solutions $\mathcal{G}_{m}^{(N)} = (\gamma_{m}^{(N)}, G_{m}^{(N)})$, where
\begin{gather*}
\gamma_m^{(N)}  =
\begin{cases}
(-1)^{\frac{m(m+1)N}{2}} \gamma_0^{(N)}, & m \ge -1, \\
(-1)^\frac{(m-2) (m-1) m (m+1)}{24} \gamma_0^{(N)}, & -2 \ge m \ge -2N-1, N:\text{even}, \\
(-1)^\frac{m (m+1) (m+2) (m+3)}{24} \gamma_0^{(N)}, & -2 \ge m \ge -2N-1, N:\text{odd}, \\
(-1)^{\frac{N(N-1)}{2}} \gamma_0^{(N)}, & m \le -2N-2,
\end{cases}\\
G_m^{(N)}  =
\begin{cases}
\dfrac{mN(N-1)}{2} Q + G_0^{(N)}, & m \ge -1, \vspace{2mm}\\
\dfrac{mN(N-1)}{2} Q + \dfrac{(m-2) m (2m+1)}{24} Q + G_0^{(N)}, & m = -2,-4,\dots,-2N, \vspace{2mm} \\
\dfrac{mN(N-1)}{2} Q + \dfrac{(m-1) (m+1) (2m-3)}{24} Q + G_0^{(N)},\!\! & m = -3,-5,\dots,-2N{-}1, \vspace{2mm}\\
-\dfrac{m N (m+N)}{2} Q - \dfrac{N (N-1) (4N+1)}{6} Q + G_0^{(N)}, & m \le -2N -2.
\end{cases}
\end{gather*}
Since $G_m^{(N)} \to \infty$ as $m \to -\infty$ in the same way as the uBi function, we call these solutions the Bi-function-type solutions.

By substituting these solutions into \eqref{eq:udCH1+} and \eqref{eq:udCH2+}, we obtain special solutions of udPII,
\begin{gather}
\mathcal{Z}_m^{(N)} = \left(\zeta_m^{(N)}, Z_m^{(N)} \right) =
\begin{cases}
\left( (-1)^{m-1}, 0 \right), & m \ge -2N-1, \\
\left( +1, -(m+2N+1) Q \right), & m \le -2N-2.
\end{cases}\label{eq:Psol_bi+}
\end{gather}
We notice that \eqref{eq:Psol_ai+} and \eqref{eq:Psol_bi+} have dif\/ferent asymptotic behavior in $Z_m^{(N)}$ for $m \to -\infty$ and in the phases for $m > 0$. Furthermore, we remark that \eqref{eq:Psol_bi+} have $N$-dependence.

Thirdly, we study Ai-function-type solutions for $N < 0$. We f\/ind that $\mathcal{G}_{m}^{(0)} = (+1,0)$ and $\mathcal{G}_{m}^{(-1)} = \mathrm{uAi} (m-1)$ solve \eqref{eq:bilinearG1-imp} and \eqref{eq:bilinearG2-imp} with $N=-1$. Starting from these simple solutions, we inductively f\/ind the solutions
$\mathcal{G}_{m}^{(N)} = (\gamma_{m}^{(N)}, G_{m}^{(N)})$ for $N=0, -1, -2, \dots$, where
\begin{gather*}
\gamma_{m}^{(N)}  =
\begin{cases}
(-1)^{\frac{(m+N)(m+N-1)N}{2}} \gamma_{1}^{(N)}, & m \ge -N, \\
\gamma_{1}^{(N)}, & m \le -N,
\end{cases} \\
G_{m}^{(N)} =
\begin{cases}
\dfrac{N(N+1)(m+N-1)}{2} Q + G_{1}^{(N)}, & m \ge -N, \vspace{2mm}\\
\dfrac{-N(m+N-1)(m-1)}{2} Q + G_{1}^{(N)}, & m \le -N.
\end{cases}
\end{gather*}

\begin{figure}[htbp]
\centering
$(a)$\includegraphics[width=.38\linewidth]{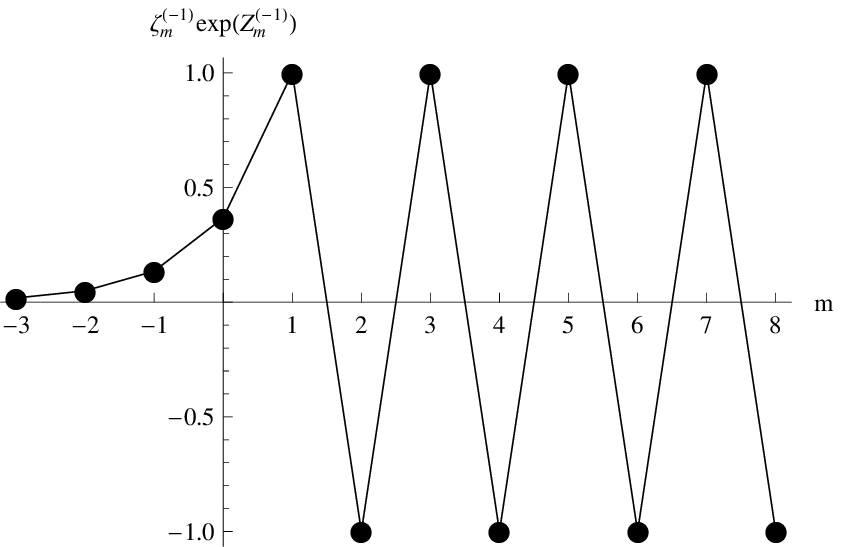}\qquad\qquad
$(b)$\includegraphics[width=.38\linewidth]{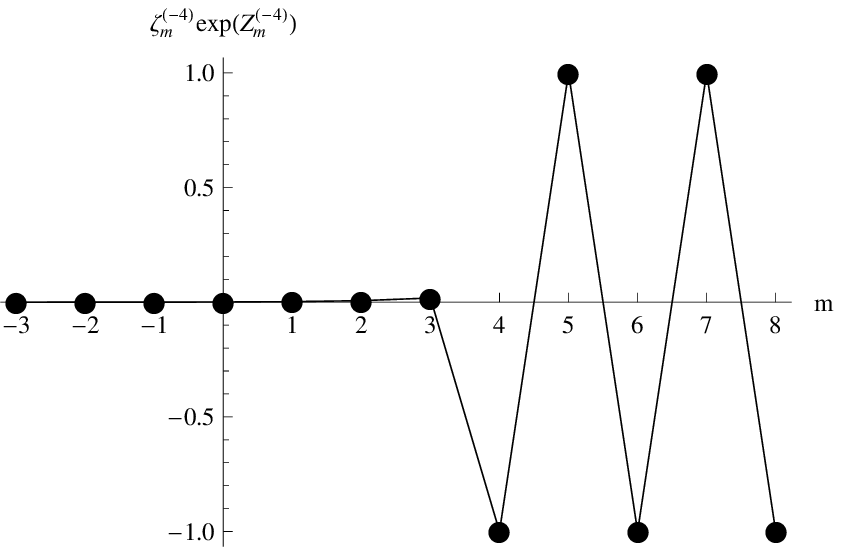}
\caption{Behavior of special solutions \eqref{eq:Psol_ai-} with $Q=-1$. $(a)$ and $(b)$ are for $N=-1$ and $N=-4$, respectively.}
\label{fig1}
\end{figure}

Substituting these $\mathcal{G}_{m}^{(N)}$ into \eqref{eq:udCH1-} and \eqref{eq:udCH2-}, we have special solutions of udPII with $A=(2N+1)Q$ for $N=-1, -2, \dots$,
\begin{gather}
\mathcal{Z}_m^{(N)} = \left(\zeta_m^{(N)}, Z_m^{(N)} \right) =
\begin{cases}
\left( (-1)^{m-1}, 0 \right), & m \ge -N, \\
\left( +1, -(m+2N+1)Q  \right), & m \le -N-1.
\end{cases} \label{eq:Psol_ai-}
\end{gather}
Typical behavior of these solutions is shown in Fig.~\ref{fig1}. They converge to 0 as $m \to -\infty$ and oscillate for $m \ge -N$. It is interesting to note that \eqref{eq:Psol_ai-} constructed from the uAi function has essentially the same structure as~\eqref{eq:Psol_bi+} constructed from the uBi function.

\begin{figure}[htbp]
\centering
$(a)$\includegraphics[width=.38\linewidth]{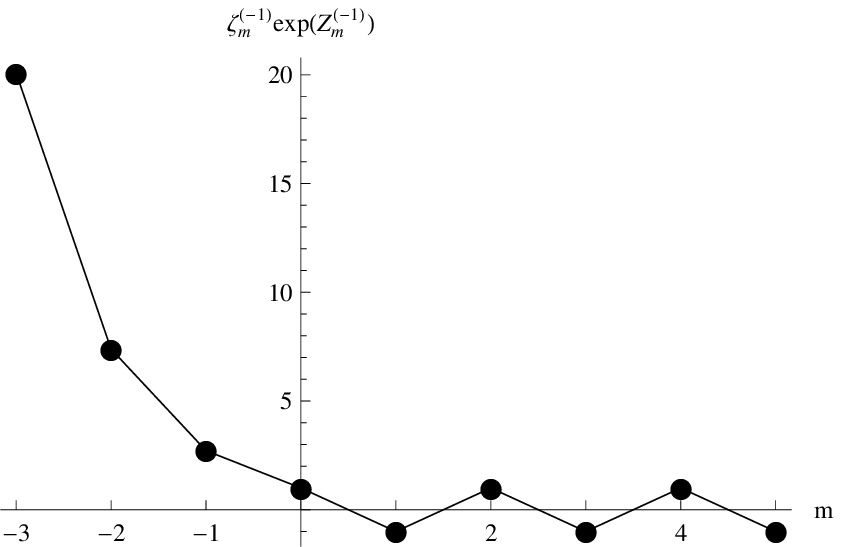}\qquad\qquad
$(b)$\includegraphics[width=.38\linewidth]{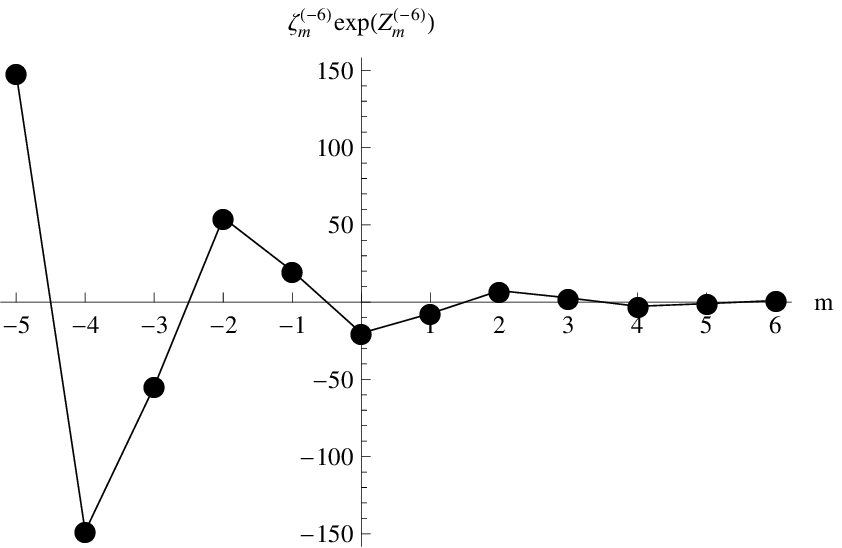}

\caption{Behavior of special solutions \eqref{eq:Psol_bi-} with $Q=-1$. $(a)$ and $(b)$ are for $N=-1$ and $N=-6$, respectively.}
\label{fig2}
\end{figure}

Finally, we study Bi-function-type solutions for $N < 0$. We f\/ind that $\mathcal{G}_{m}^{(0)} = (+1,0)$ and $\mathcal{G}_{m}^{(-1)} = \mathrm{uBi} (m-1)$ solve \eqref{eq:bilinearG1-imp} and \eqref{eq:bilinearG2-imp} with $N=-1$. We construct solutions $\mathcal{G}_{m}^{(N)}$ of the equations with $N \le -2$ for a given function $\mathcal{G}_{m}^{(N+1)}$ and assigned values of $\mathcal{G}_{1}^{(N)}$ and $\mathcal{G}_{2}^{(N)}$. We further assume that $\mathcal{G}_{1}^{(N)}$ and $\mathcal{G}_{2}^{(N)}$ are chosen so that $\mathcal{G}_{m}^{(N)}$  for $m \le 0$ are uniquely determined in \eqref{eq:bilinearG1-imp} and \eqref{eq:bilinearG2-imp}. We again inductively obtain the solutions $\mathcal{G}_{m}^{(N)} = (\gamma_{m}^{(N)}, G_{m}^{(N)})$, where
\begin{gather*}
\gamma_m^{(N)} =
\begin{cases}
(-1)^{\frac{N(N+2)}{8}} \gamma_1^{(N)}, & m \ge 3-N,N:\text{even}, \vspace{1mm}\\
(-1)^{\frac{(m+N-3)(m+N-4)}{2}+\frac{(N-1)(N-3)}{8}+1} \gamma_1^{(N)}, & m \ge 3-N,N:\text{odd}, \\
(-1)^{p_{N,m}} \gamma_1^{(N)}, & N-1 \le m \le 2-N, \\
(-1)^{\frac{(N+2)(N+3)(N+4)(N+5)}{24}+1} \gamma_1^{(N)}, & m \le N-2,
\end{cases} \\
p_{N,m} = \dfrac{(m-N-3)(m-N-4)(m-N-5)(m-N-6)}{24} \\
\phantom{p_{N,m} =}{} + \dfrac{(N+2)(N+3)(N+4)(N+5)}{24},
\\
G_m^{(N)} =
\begin{cases}
\dfrac{(m-1) N(N+1)}{2} Q - \dfrac{N (N+2) (2N-1)}{24} Q + G_1^{(N)}, \vspace{1mm}\\
\hspace*{60mm}m \ge 3-N, N:\text{even},\vspace{1mm}\\
\dfrac{(m-1) N(N+1)}{2} Q - \dfrac{(N-1) (N+1) (2N+3)}{24} Q + G_1^{(N)}, \vspace{1mm}\\
\hspace*{60mm} m \ge 3-N, N:\text{odd},   \vspace{1mm}\\
\dfrac{(m-1)N(m+3N+2)}{4} Q + \dfrac{(m-2) m (2m+1)}{24} Q + G_1^{(N)},\vspace{1mm}\\
\hspace*{60mm} N-1 \le m \le 2-N, m:\text{even}, N:\text{even}, \vspace{1mm}\\
\dfrac{(m-1)N(m+3N+2)}{4} Q + \dfrac{(m-2) m (2m+1)+6}{24} Q + G_1^{(N)},\vspace{1mm}\\
\hspace*{60mm} N-1 \le m \le 2-N, m:\text{even}, N:\text{odd}, \vspace{1mm}\\
\dfrac{(m-1)N(m+3N+2)}{4} Q + \dfrac{(m-1) (m+1) (2m-3)}{24} Q + G_1^{(N)}, \vspace{1mm}\\
\hspace*{60mm} N-1 \le m \le 2-N, m:\text{odd}, \vspace{1mm}\\
\dfrac{m N (m+N)}{2} Q + \dfrac{N (2N^2-15N-14)}{24} Q + G_1^{(N)},\vspace{1mm}\\
\hspace*{60mm} m \le N -2, N:\text{even},\vspace{1mm}\\
\dfrac{m N (m+N)}{2} Q + \dfrac{(N+1) (2N^2-17N+3)}{24} Q + G_1^{(N)},\vspace{1mm}\\
\hspace*{60mm} m \le N -2, N:\text{odd}.
\end{cases}
\end{gather*}
From these solutions, we have special solutions of udPII,
\begin{gather}
\mathcal{Z}_m^{(N)} = \big(\zeta_m^{(N)},\ Z_m^{(N)} \big) \nonumber\\
\phantom{\mathcal{Z}_m^{(N)}}{} =
\begin{cases}
\left( (-1)^{m}, 0 \right), & m \ge -N-1, \\
\left( (-1)^{\frac{m-N}{2}}, \dfrac{m+N}{2} Q \right), & |m| \le -N-2, m-N:\text{even}, \\
\left( (-1)^{\frac{m-N-1}{2}}, \dfrac{m+N+1}{2} Q \right), & |m| \le -N-2, m-N:\text{odd}, \\\
\left( +1, m Q \right), & m \le N+1.
\end{cases}\label{eq:Psol_bi-}
\end{gather}
Typical behavior of these solutions is shown in Fig.~\ref{fig2}. Note that \eqref{eq:Psol_ai-} and \eqref{eq:Psol_bi-} have dif\/ferent asymptotic behavior in $Z_m^{(N)}$ as $m \to -\infty$ and in the phases for  $m \ge -N$.
We also comment that, although \eqref{eq:Psol_bi-} is similar to~\eqref{eq:Psol_ai+},~\eqref{eq:Psol_bi-} has more complicated internal structure.

\section{Concluding remarks}\label{section4}

In this paper we have presented a class of special solutions for the $p$-ultradiscrete analog of $q$-PII. The solutions are classif\/ied into four
groups; Ai-function-type and Bi-function-type solutions for the system
parameter $N \ge 0$, and those for $N < 0$. In the preceding paper~\cite{IKMMS} are
given only the Ai-function-type solutions for $N \ge 0$, which do not depend on
$N$. Three other groups which are newly given in this paper do depend on $N$.
Moreover, the solutions of each group have dif\/ferent structures. For example,
we observe dif\/ferences between the Ai- and Bi-function-type solutions in
their asymptotic amplitude and phases, which may ref\/lect the structure of
solutions of dif\/ference and continuous equations. The Bi-function-type
solutions for $N < 0$ have fairly complicated internal structure,
although we do not know the origin of these structures yet. At any rate,
these results may indicate the richness of solution space of the ultradiscrete
equation.

For the continuous and discrete Airy equations, linear combination of Ai and
Bi functions give their general solutions. In the ultradiscrete case, $\max (f,g)$
corresponds to the linear combination of functions $f$ and $g$. Hence, we
believe that the cases we treated in this paper cover quite wide class of
special solutions of the ultradiscrete equations.

Our method of constructing solutions is intuitive and purely based on the
ultradiscrete equations. We believe that the solutions we obtain correspond
to those of $q$-PII represented by the Casorati determinant of size $|N|$
whose elements are given by the $q$-dif\/ference Ai or Bi function. It is a
future problem to clarify the relationship between discrete and ultradiscrete
solutions through a limiting procedure. It is also a future problem to
construct $p$-ultradiscrete analogs of other Painlev\'{e} equations and their
special solutions.

\subsection*{Acknowledgements}
This work was supported by JSPS KAKENHI 21760063 and 21560069.

\pdfbookmark[1]{References}{ref}
\LastPageEnding

\end{document}